\documentclass[9pt,conference]{IEEEtran}

\usepackage[preprint]{waspaa25}

\usepackage{cite}
\usepackage{amsmath,amssymb,amsfonts}
\usepackage{algorithmic}
\usepackage{graphicx}
\usepackage{textcomp}
\usepackage{xcolor}
\usepackage{scalerel}
\usepackage[capitalize,nameinlink]{cleveref}
\usepackage{xifthen}
\usepackage{xspace}
\usepackage{xfrac}
\usepackage{multicol}
\usepackage[inline]{enumitem}
\usepackage{booktabs}
\usepackage{float}
\usepackage{siunitx}
\usepackage{tikz}
\usepackage{bm}
\usepackage{url}
\usepackage{pifont}
\usepackage{microtype}
\usepackage{subcaption}
\usepackage{newtxmath}
\usepackage{anyfontsize}
\usetikzlibrary{shapes,arrows,positioning,perspective,fit,3d,backgrounds,decorations.pathreplacing,calc,arrows.meta,colorbrewer}
\newcommand{\ufj}{\upsilon_j}
\newcommand{\uftrgt}{\upsilon_{\text{trgt}}}

\DeclareMathOperator*{\argmax}{arg\,max}

\DeclareMathOperator*{\softmax}{softmax}

\DeclareMathOperator*{\upsample}{upsample}

\title{Adaptive Slimming for Scalable and Efficient Speech Enhancement\\
\thanks{This work has received funding from the European Union’s Horizon research and innovation program under grant agreement No 101070374.}
}

\name{
Riccardo Miccini$^{\flat \natural \sharp}$,
Minje Kim$^{\natural}$,
Clément Laroche$^{\flat}$, 
Luca Pezzarossa$^{\sharp}$,
Paris Smaragdis$^{\natural}$
\vspace{-0.3em}
}

\address{
$^{\flat}$GN Audio, Denmark; \;
$^{\natural}$University of Illinois at Urbana-Champaign, USA; \;
$^{\sharp}$Technical University of Denmark, Denmark
}

\begin{document}
\bstctlcite{MyBSTcontrol}

\maketitle

\begin{abstract}
Speech enhancement (SE) enables robust speech recognition, real-time communication, hearing aids, and other applications where speech quality is crucial.
However, deploying such systems on resource-constrained devices involves choosing a static trade-off between performance and computational efficiency. 
In this paper, we introduce dynamic slimming to DEMUCS, a popular SE architecture, making it scalable and input-adaptive. 
Slimming lets the model operate at different utilization factors (UF), each corresponding to a different performance/efficiency trade-off, effectively mimicking multiple model sizes without the extra storage costs. 
In addition, a router subnet, trained end-to-end with the backbone, determines the optimal UF for the current input. 
Thus, the system saves resources by adaptively selecting smaller UFs when additional complexity is unnecessary. 
We show that our solution is Pareto-optimal against individual UFs, confirming the benefits of dynamic routing.
When training the proposed dynamically-slimmable model to use \qty{10}{\percent} of its capacity on average, we obtain the same or better speech quality as the equivalent static \qty{25}{\percent} utilization while reducing MACs by \qty{29}{\percent}.\footnote{Samples at \url{https://miccio-dk.github.io/adaptive-slimming-speech-enh/}} 
\end{abstract}


\section{Introduction}
\label{sec:intro}

Speech enhancement (SE) systems are a crucial part of telecommunication, teleconferencing, and assistive technology, improving remote collaboration, user experience, and quality of life.
In recent years, SE solutions based on deep learning made significant progress, thanks to their ability to learn from vast amounts of data~\cite{braun_data_2020,fedorov_tinylstms_2020,shankar_real-time_2020,braun_towards_2021,yu_pfrnet_2022,rong_gtcrn_2024}.
However, deploying these systems on resource-constrained embedded devices such as headsets, speakerphones, or hearing aids requires a careful trade-off between denoising performance and computational needs.
This can be challenging to achieve, given the broad range of acoustic conditions and noise levels experienced in the real world.
Because devices must handle rare but challenging acoustic scenarios without failing, they require sufficiently large models trained from a large dataset for worst-case conditions. 
Consequently, in typical scenarios that involve relatively clean inputs, SE models are over-provisioned, wasting computational resources and energy.

Dynamic neural networks (DynNN) offer a solution to this problem, providing specialized architectures and training strategies to develop artificial neural networks that can manipulate certain aspects of their computational graph at inference time, often based on the input~\cite{han_dynamic_2022}.
DynNNs with scalable depth, width, or adaptive/conditional connections have been successfully applied to computer vision~\cite{guo_dynamic_2019,yu_slimmable_2019,herrmann_channel_2020,gao_dynamic_2019,li_dynamic_2021} and, as of recently, are being investigated on audio-processing tasks~\cite{chen_dont_2021,bralios_latent_2023,miccini_scalable_2025,elminshawi_dynamic_2024}.
In this paper, we introduce \textit{dynamic slimming} to DEMUCS~\cite{defossez_real_2020}, a popular SE architecture, to make it scalable and input-adaptive.
Slimming is a form of width-wise dynamism that lets the model operate at different \textit{utilization factors} (UF), corresponding to discrete subsets of weights, each providing a different performance/efficiency trade-off.
Intuitively, higher UFs correspond to larger networks, resulting in better denoising performance, but at higher computational costs.
This is conceptually equivalent to deploying multiple models of different sizes without incurring the cost of storing all their weights, such as a sparse mixture of local experts for SE~\cite{sivaraman_sparse_2020}, thanks to its structured architecture.

\begin{figure}[t]
    \centering
    \centerline{\includegraphics[width=1.0\columnwidth]{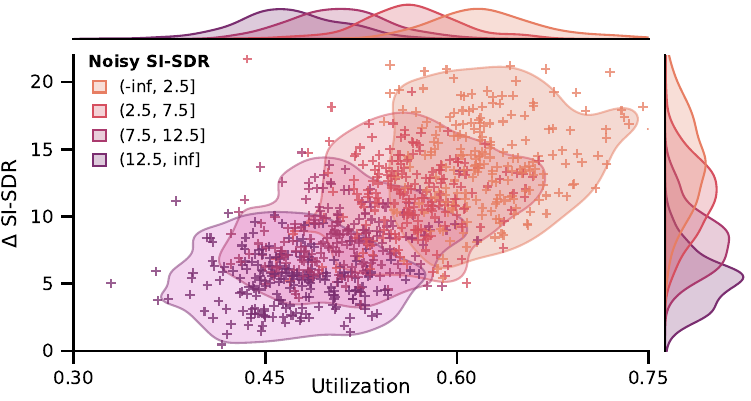}}
    \vspace{-0.5em}
    \caption{Relationship between average utilization and SI-SDR improvement for model trained with $\uftrgt = 0.5$, showing how our proposed solution processes noisier input data (lighter points) using more resources, causing a larger improvement. 
    }
    \label{fig:dynslim_jointplot}
\end{figure}

We hypothesize that the UF concept is associated with the difficulty of the SE task.
Input noise conditions can range from easy (e.g., high SNR or static noise) to more challenging (e.g., low SNR, non-stationary, or speech-like noise). 
Thus, using a large model across a range of SE tasks can induce redundancy in its architecture, wasting resources when the input is trivial. 
In our work, a routing module estimates the optimal UF for each utterance based on the input characteristics and a desired average utilization target, ensuring an advantageous trade-off between computational efficiency and speech quality by selecting smaller UFs when larger ones would offer limited additional benefits (as demonstrated in \cref{fig:dynslim_jointplot}).
The resulting system is an adaptively-slimmable neural network that can match the denoising performance of a given static baseline while using fewer computational resources on average. 
We also observe that dynamic routing is Pareto-optimal when compared with statically selecting a specific UF because, by adapting to the input, the proposed system selects the most advantageous trade-off points, reducing the average computational costs. 
We contribute the following:
\begin{enumerate*}[label={\arabic*)}]
    \item Performance and efficiency improvements on the DEMUCS baseline and architectural adaptations to support multiple performance/efficiency trade-offs through slimming;
    \item Design of a lightweight routing sub-network for input-dependent scalability, along with training strategy;
    \item Model evaluation with respect to its training hyperparameters.
\end{enumerate*}

\begin{figure*}[t]
    \centering
    \resizebox{0.97\textwidth}{!}{\begin{tikzpicture}[
    node distance=2.5em,
    auto,
]

\tikzset{every node/.append style={transform shape, align=center, text centered}}
\tikzset{box/.style={rectangle, draw, thin, fill=white}}

\tikzset{boxenc/.style={box, minimum width=10em, minimum height=2em, inner sep=0.25em, trapezium, trapezium angle=-80, trapezium stretches body, shape border rotate=90, fill=Greys-F}}
\tikzset{boxdec/.style={boxenc, trapezium angle=80}}
\tikzset{boxgru/.style={box, minimum height=1em, minimum width=5em, inner sep=0em, fill=Paired-G}}

\tikzset{boxsub/.style={box, minimum height=5em, minimum width=0.8em, inner sep=0em, font=\tiny, fill=Greys-D}}
\tikzset{boxsubh/.style={boxsub, minimum width=4.5em, minimum height=0.8em,fill=Paired-E}}

\tikzset{boxres/.style={boxsub, minimum height=3em, rounded corners=2pt, fill=white}}
\tikzset{boxresh/.style={boxres, minimum width=3em, minimum height=0.8em}}

\tikzset{boxdots/.style={minimum height=1em, minimum width=2.5em, font=\scriptsize, inner sep=0em,fill=none}}
\tikzset{inout/.style={inner sep=0.25em}}

\tikzset{groupgru/.style={box, inner sep=0.25em, fill=Paired-H!80}}
\tikzset{groupufp/.style={box, inner sep=0.25em, fill=Paired-F!80}}

\tikzset{ar/.style={draw, thin, rounded corners, -{to[scale=0.7]}}}
\tikzset{arufp1/.style={ar, densely dotted, draw=black!65}}
\tikzset{arufp2/.style={arufp1, -{Circle[open,scale=0.7]}}}
\tikzset{arufpw/.style={ar, thick, draw=white, -}}

\node [boxgru] (rnn1) {$\scriptstyle\text{GRU}_1$};
\node [boxgru,  below=0.5em of rnn1] (rnn2) {$\scriptstyle\text{{GRU}}_2$};
\node [boxdots, below=0em of rnn2] (rnn3) {$\boldsymbol{\cdots}$};
\node [boxgru,  below=0em of rnn3] (rnng) {$\scriptstyle\text{GRU}_M$};

\path let \p1 = (rnn1.north west), \p2 = (rnng.south) in
        node[boxsub, anchor=north east, fill=Paired-G] at (\x1-0.3em, \y1) [minimum height={\y1-\y2-0.5pt}] (rnnsplit) {\rotatebox{90}{$\text{Split}$}};
\path let \p1 = (rnn1.north east), \p2 = (rnng.south) in
        node[boxsub, anchor=north west, fill=Paired-G] at (\x1+0.3em, \y1) [minimum height={\y1-\y2-0.5pt}] (rnnconcat) {\rotatebox{90}{$\text{Concatenate}$}};

\begin{scope}[on background layer]
    \node [groupgru, fit={(rnnsplit) (rnn1) (rnng) (rnnconcat)}] (rnn) {};
\end{scope}

\node [boxenc,  left=of rnn, minimum width=6em] (encl) {$\mathcal{E}_L$};
\node [boxdots, left=0em of encl] (encdots) {$\boldsymbol{\cdots}$};
\node [boxenc,  left=0em of encdots, minimum width=7.3em] (enc3) {$\mathcal{E}_3$};

\node [boxsub, left=2.75em of enc3, fill=none, draw=none, minimum height=0.8em] (encspl) {};
\node[fill=black, circle, minimum size=2pt, inner sep=0pt] (encdot) at (encspl.center) {};
\node [boxsub, left=0.25em of encspl] (encglu) {\rotatebox{90}{GLU}};
\node [boxsub, left=0.25em of encglu, fill=Paired-B!50!Paired-A] (encpw) {\rotatebox{90}{PWConv1D}};
\node [boxsub, left=0.25em of encpw] (encrelu) {\rotatebox{90}{ReLU}};
\node [boxsub, left=0.25em of encrelu, fill=Paired-D!50!Paired-C] (encc) {\rotatebox{90}{Conv1D}};

\begin{scope}[on background layer]
    \node [boxenc,  minimum width=9.8em, fit={(encspl) (encc)}] (enc2) {};
    \node[below=0.5em of enc2.right side, inner sep=0em] (dec2l) {$\mathcal{E}_2$};
\end{scope}
\node [boxenc,  left=of enc2, minimum width=11.2em] (enc1) {$\mathcal{E}_1$};

\node [boxres, left=2em of enc1] (inup) {\rotatebox{90}{Upsample}};
\node [fill=black, circle, minimum size=2pt, inner sep=0pt, left=4em of inup] (indot) {};
\node [inout, left=4em of indot] (in) {\strut$x$};

\node [boxdec,  right=of rnn, minimum width=6em] (decl) {$\mathcal{D}_L$};
\node [boxdots, right=0em of decl] (decdots) {$\boldsymbol{\cdots}$};
\node [boxdec,  right=0em of decdots, minimum width=7.3em] (dec3) {$\mathcal{D}_3$};

\node [boxsub, right=2.75em of dec3, circle, minimum height=0.8em] (decsum) {\rotatebox{90}{$+$}};
\node [boxsub, right=0.25em of decsum, fill=Paired-B!50!Paired-A] (decpw) {\rotatebox{90}{PWConv1D}};
\node [boxsub, right=0.25em of decpw] (decglu) {\rotatebox{90}{GLU}};
\node [boxsub, right=0.25em of decglu, fill=Paired-D!50!Paired-C] (dectr) {\rotatebox{90}{TranspConv1D}};
\node [boxsub, right=0.25em of dectr] (decrelu) {\rotatebox{90}{ReLU}};

\begin{scope}[on background layer]
    \node [boxdec,  minimum width=9.8em, fit={(decsum) (decrelu)}] (dec2) {};
    \node[below=0.5em of dec2.right side, inner sep=0em] (dec2l) {$\mathcal{D}_2$};
\end{scope}
\node [boxdec,  right=of dec2, minimum width=11.2em] (dec1) {$\mathcal{D}_1$};

\node [boxres, right=2em of dec1, minimum height=4em] (outdo) {\rotatebox{90}{Downsample}};
\node [inout, right=3em of outdo] (out) {\strut$\hat{s}$};

\node [boxsubh, above=1.25em of indot.center] (ufpconv) {Conv1D};
\node [boxsubh, above=0.25em of ufpconv] (ufprelu) {ReLU};
\node [boxsubh, above=0.25em of ufprelu] (ufpgru) {DiagonalGRU};
\node [boxsubh, above=0.25em of ufpgru] (ufppwconv) {PWConv1D};
\node [boxsubh, above=0.25em of ufppwconv] (ufpsoftm) {GumbelSoftmax};

\begin{scope}[on background layer]
    \node [groupufp, fit={(ufpconv) (ufpsoftm)}] (ufp) {};
    \node[right=0.25em of ufp, inner sep=0em] (ufpl) {$\mathcal{R}$};
\end{scope}

\node[boxresh, above=1.5em of ufp] (ufpres) {Resample};
\foreach \n in {enc1, enc2, enc3, dec3, dec2, dec1} {
    \node[boxresh] (\n-ups) at (\n |- ufpres) {Upsample};
}

\node[below=0.25em of indot.center, inner sep=0.25em] (ufpl) {\footnotesize$\left[1, T\right]$};
\node[above=5.1em of rnn, inner sep=0.25em] {\footnotesize$\left[J, \frac{UT}{S^L}\right]$};

\draw [ar] (in) -- (inup);
\draw [ar] (inup) -- (enc1) node[midway, above, inner sep=0.5em] {\rotatebox{90}{\footnotesize$\left[1, UT\right]$}};
\draw [ar] (enc1) -- (enc2) node[midway, above, inner sep=0.5em] {\rotatebox{90}{\footnotesize$\left[H, \frac{UT}{S}\right]$}};
\draw [ar,-] (encglu) -- (enc2);
\draw [ar] (enc2) -- (enc3) node[midway, above, inner sep=0.5em] {\rotatebox{90}{\footnotesize$\left[2H, \frac{UT}{S^2}\right]$}};
\draw [ar] (encl) -- (rnn) node[midway, above, inner sep=0.5em] {\rotatebox{90}{\footnotesize$\left[2^{L-1}H, \frac{UT}{S^L}\right]$}};
\draw [ar,-,densely dashed] (enc2.south) |- ($(encspl.south) + (0em, -3em)$) -- (encdot);

\draw [ar] (rnn) -- (decl) node[midway, above, inner sep=0.5em] {\rotatebox{90}{\footnotesize$\left[2^{L-1}H, \frac{UT}{S^L}\right]$}};
\draw [ar] (dec3) -- (dec2) node[midway, above, inner sep=0.5em] {\rotatebox{90}{\footnotesize$\left[2H, \frac{UT}{S^2}\right]$}};
\draw [ar] (dec2) -- (dec1) node[midway, above, inner sep=0.5em] {\rotatebox{90}{\footnotesize$\left[H, \frac{UT}{S}\right]$}};
\draw [ar] (dec1) -- (outdo) node[midway, above, inner sep=0.5em] {\rotatebox{90}{\footnotesize$\left[1, UT\right]$}};
\draw [ar] (outdo) -- (out) node[midway, below, inner sep=0.5em] {\footnotesize$\left[1, T\right]$};
\draw [ar,densely dashed] (dec2.south) |- ($(decsum.south) + (0em, -3em)$) -- (decsum);

\draw [ar] (encl.south) |- ($(rnn.south) + (0em, -1em)$) -| (decl.south);
\draw [ar] (enc3.south) |- ($(rnn.south) + (0em, -2em)$) -| (dec3.south);
\draw [ar,-] (enc2.south) |- ($(rnn.south) + (0em, -3em)$) -| (dec2.south);
\draw [ar] (enc1.south) |- ($(rnn.south) + (0em, -4em)$) -| (dec1.south);

\draw [ar] (indot) -- (ufp.south);
\draw [arufp1] (ufp.north) -- (ufpres.south) node[midway, right, inner sep=0.5em] {\footnotesize$\left[J, T^{\ast}\right]$};

\begin{scope}[on background layer]
    \foreach \n in {dec1, dec2, dec3, decl, encl, enc3, enc2, enc1} {
        \ifthenelse{\NOT \equal{\n}{decl} \AND \NOT \equal{\n}{encl}}{
            \draw [arufpw] (ufpres.north) -- ++(0em, 1em) -| (\n-ups.north);
            \draw [arufp1] (ufpres.north) -- ++(0em, 1em) -| (\n-ups.north);
            \draw [arufp2] (\n-ups.south) -- (\n.north);
        }{
            \draw [arufpw] (ufpres.north) -- ++(0em, 1em) -| (\n.north);
            \draw [arufp2] (ufpres.north) -- ++(0em, 1em) -| (\n.north); 
        }
    }
\end{scope}

\end{tikzpicture}}
    \caption{Overall solution showing DEMUCS backbone with encoder $\mathcal{E}_i$ and decoder $\mathcal{D}_i$ blocks (gray), the grouped GRU bottleneck (orange), and routing subnet $\mathcal{R}$ (red). Connections between blocks are annotated with signal dimensionality. Dotted connections represent UF arguments.}
    \label{fig:dynslim_demucs}
\end{figure*}
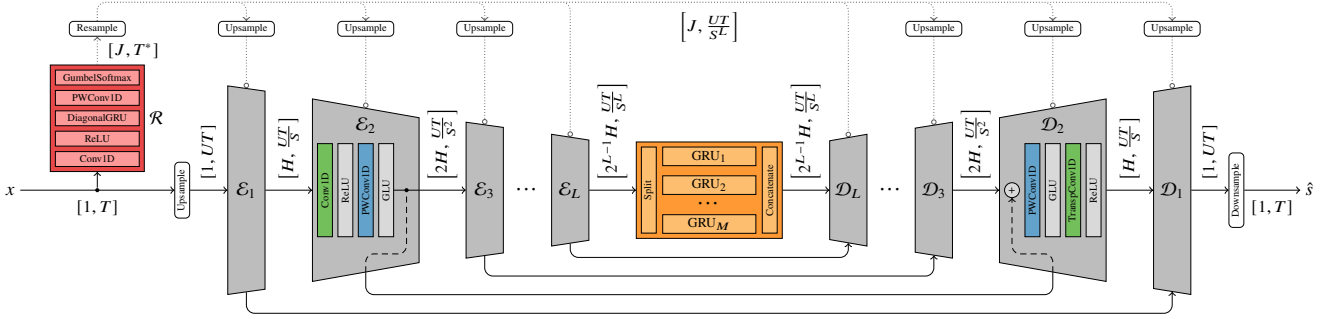


\begin{figure}[t]
    \centering
    \begin{subfigure}[t]{0.47\linewidth}
        \centering
        \resizebox{!}{4.5cm}{
            \begin{tikzpicture}[
    x=1em,y=1em,z=-0.4em,
    node distance=2em,
    auto,
]

\tikzset{every node/.append style={transform shape, align=center, text centered}}
\tikzset{box/.style={rectangle, draw, thin, fill=Greys-D, minimum width=8em}}

\tikzset{boxc/.style={box, inner sep=0.5em}}
\tikzset{boxa/.style={box, inner sep=0.5em}}
\tikzset{dot/.style={fill=black, circle, minimum size=2pt, inner sep=0pt}}
\tikzset{inout/.style={inner sep=0.25em, minimum width=4em, fill=none}}
\tikzset{ar/.style={draw, thin, -{to[scale=0.7]}}}
\tikzset{arlbl/.style={midway, left, inner sep=0.5em}}
\tikzset{lbrace/.style={decorate,decoration={brace,mirror,raise=2pt,aspect=#1}}}
\tikzset{lfbrace/.style={decorate,decoration={brace,mirror,raise=5pt,aspect=#1}}}

\node [inout] (in) {$\mathcal{E}_{i-1}\strut$};
\node [boxc, below=2em of in,fill=Paired-D!50!Paired-C] (conv) {Conv1D};
\node [boxa, below=0.75em of conv] (relu) {ReLU};
\node [boxc, below=2em of relu,fill=Paired-B!50!Paired-A] (pwconv) {PWConv1D};
\node [boxa, below=2em of pwconv] (glu) {GLU};
\node [dot, below=2em of glu] (dot) {};
\node [inout, below left=1em and 0em of dot] (oute) {$\mathcal{E}_{i+1}\strut$};
\node [inout, below right=1em and 0em of dot] (outd) {$\mathcal{D}_{i}\strut$};

\node [fill=none, right=3.2em of conv, minimum width=2.5em, minimum height=4.5em] (convcanv) {};
\node [fill=none] (tensorlbl) at (convcanv |- in) {Slimmed weights};
\pgfmathsetmacro{\cubex}{2}
\pgfmathsetmacro{\cubey}{4}
\pgfmathsetmacro{\cubez}{1.2}
\pgfmathsetmacro{\cubefr}{0.75}
\begin{scope}[shift={($(convcanv.north) + (0,-0.5em)$)},line join=round]
    \draw[fill=Paired-D] let \p1 = (0,0,\cubez) in
    (0,0,0) -- ++(-0.5*\cubex,0,0) -- ++(0.5*\x1,0) coordinate (nw) -- ++(0,-\cubey,0) coordinate (sw) -- ++(\cubex,0,0) node [midway,below,inner sep=0.25em] {\scriptsize$C$} -- ++(0,\cubey,0) coordinate (ne) -- cycle
    (ne) -- ++(0,0,-\cubez) -- ++(0,-\cubey,0) coordinate (zc) -- ++(0,0,\cubez) node [midway,below,sloped,inner sep=0.25em] {\scriptsize$K$} -- cycle
    (ne) -- ++(0,0,-\cubez) -- ++(-\cubex,0,0) -- ++(0,0,\cubez) -- cycle;
    
    \draw[fill=Paired-C] let \p1 = (0,0,\cubez) in
    (0,0,0) -- ++(-0.5*\cubex,0,0) -- ++(0.5*\x1,0) -- ++(0,-\cubefr*\cubey,0) -- ++(\cubex,0,0) -- ++(0,\cubefr*\cubey,0) coordinate (ne) -- cycle
    (ne) -- ++(0,0,-\cubez) -- ++(0,-\cubefr*\cubey,0) -- ++(0,0,\cubez) -- cycle
    (ne) -- ++(0,0,-\cubez) -- ++(-\cubex,0,0) -- ++(0,0,\cubez) -- cycle;
    
    \draw[lbrace=0.25] (nw) -- (sw) node[pos=0.25,left=4pt,inner sep=0.25em] {\rotatebox{90}{\scriptsize$2C$}};
    
    \draw[lfbrace=0.5] (nw) ++(0,-\cubefr*\cubey,0) -- (sw) node[midway,left=7pt,inner sep=0.25em] {\rotatebox{90}{\scriptsize$\left\lceil2C\ufj\right\rceil$}};
\end{scope}

\node [fill=none, right=2.2em of pwconv, minimum width=4.25em, minimum height=4.25em] (pwconvcanv) {};
\pgfmathsetmacro{\cubex}{4}
\pgfmathsetmacro{\cubey}{7.5}
\pgfmathsetmacro{\cubez}{0.5}
\pgfmathsetmacro{\cubefr}{0.75}
\pgfmathsetmacro{\cubefrr}{1-\cubefr}
\begin{scope}[shift={($(pwconvcanv.north) + (0,-0.25em)$)},line join=round]
    \draw[fill=Paired-B] let \p1 = (0,0,\cubez) in
    (0,0,0) -- ++(-0.5*\cubex,0,0) -- ++(0.5*\x1,0) -- ++(0,-\cubey,0) node[midway,left,inner sep=0.25em] {\rotatebox{90}{\scriptsize$4C$}} coordinate (sw) -- ++(\cubex,0,0) coordinate (se) -- ++(0,\cubey,0) coordinate (ne) -- cycle
    (ne) -- ++(0,0,-\cubez) -- ++(0,-\cubey,0) coordinate (zc) -- ++(0,0,\cubez) node [midway,below,sloped,inner sep=0.25em] {\scriptsize$1$} -- cycle
    (ne) -- ++(0,0,-\cubez) -- ++(-\cubex,0,0) -- ++(0,0,\cubez) -- cycle;
    
    \draw[fill=Paired-A] let \p1 = (0,0,\cubez) in
    (0,0,0) -- ++(-\cubefrr*\cubex,0,0) -- ++(0.5*\x1,0) -- ++(0,-\cubey,0) -- ++(\cubefr*\cubex,0,0) -- ++(0,\cubey,0) coordinate (ne) -- cycle
    (ne) -- ++(0,0,-\cubez) -- ++(0,-\cubey,0) -- ++(0,0,\cubez) -- cycle
    (ne) -- ++(0,0,-\cubez) -- ++(-\cubefr*\cubex,0,0) -- ++(0,0,\cubez) -- cycle;

    \draw[lbrace=0.75] (sw) -- (se) node[pos=0.75,below=4pt,inner sep=0.25em] {\scriptsize$2C$};
    
    \draw[lfbrace=0.5] (sw) -- ++(\cubefrr*\cubex,0,0) node[midway,below=7pt,inner sep=0.25em] {\scriptsize$\left\lceil2C\ufj\right\rceil$};
\end{scope}


\draw [ar] (in) -- (conv)     node[arlbl] {\scriptsize$\left[C, T\right]$};
\draw [ar] (conv) -- (relu);
\draw [ar] (relu) -- (pwconv) node[arlbl] {\scriptsize$\left[2C\ufj, \frac{T}{S}\right]$};
\draw [ar] (pwconv) -- (glu)  node[arlbl] {\scriptsize$\left[4C, \frac{T}{S}\right]$};
\draw [ar,-] (glu) -- (dot)   node[arlbl] {\scriptsize$\left[2C, \frac{T}{S}\right]$};
\draw [ar] (dot) -| (oute);
\draw [ar] (dot) -| (outd);

\pgfresetboundingbox \useasboundingbox[draw=none] (tensorlbl.north east) -| (glu.west) |- (oute) -| (tensorlbl.north east);


\end{tikzpicture}
        }
        \caption{Encoder block $\mathcal{E}_i$}
        \label{fig:slim_block_enc}
    \end{subfigure}%
    \hfill
    \begin{subfigure}[t]{0.47\linewidth}
        \centering
        \resizebox{!}{4.5cm}{
            \begin{tikzpicture}[
    x=1em,y=1em,z=-0.4em,
    node distance=2em,
    auto,
]

\tikzset{every node/.append style={transform shape, align=center, text centered}}
\tikzset{box/.style={rectangle, draw, thin, fill=Greys-D, minimum width=8em}}

\tikzset{boxc/.style={box, inner sep=0.5em}}
\tikzset{boxa/.style={box, inner sep=0.5em}}
\tikzset{dot/.style={fill=black, circle, minimum size=2pt, inner sep=0pt}}
\tikzset{inout/.style={inner sep=0.25em, minimum width=4em, fill=none}}
\tikzset{ar/.style={draw, thin, -{to[scale=0.7]}}}
\tikzset{arlbl/.style={midway, left, inner sep=0.5em}}
\tikzset{lbrace/.style={decorate,decoration={brace,mirror,raise=2pt,aspect=#1}}}
\tikzset{lfbrace/.style={decorate,decoration={brace,mirror,raise=5pt,aspect=#1}}}

\node [draw, circle, inner sep=0.1em, fill=Greys-D] (sum) {$+$};
\node [inout, above left=1em and 0em of sum.center] (in1) {$\mathcal{E}_i\strut$};
\node [inout, above right=1em and 0em of sum.center] (in2) {$\mathcal{D}_{i+1}\strut$};
\node [boxc, below=2em of sum.center,fill=Paired-B!50!Paired-A] (pwconv) {PWConv1D};
\node [boxa, below=2em of pwconv] (glu) {GLU};
\node [boxc, below=2em of glu,fill=Paired-D!50!Paired-C] (trconv) {TranspConv1D};
\node [boxa, below=0.75em of trconv] (relu) {ReLU};
\node [inout, below=2em of relu] (out) {$\mathcal{D}_{i-1}\strut$};

\node [fill=none, right=3em of pwconv, minimum width=4.25em, minimum height=4.25em] (pwconvcanv) {};
\node [fill=none] (tensorlbl) at (pwconvcanv |- in2) {Slimmed weights};
\pgfmathsetmacro{\cubex}{4}
\pgfmathsetmacro{\cubey}{7.5}
\pgfmathsetmacro{\cubez}{0.5}
\pgfmathsetmacro{\cubeuf}{0.25}         
\pgfmathsetmacro{\cubehuf}{0.5*\cubeuf} 
\pgfmathsetmacro{\cubehy}{0.5*\cubey}   
\pgfmathsetmacro{\cubeufy}{\cubehy*\cubeuf} 
\begin{scope}[shift={($(pwconvcanv.north) + (0,1em)$)},line join=round]
    \draw[fill=Paired-A] let \p1 = (0,0,\cubez) in
    (0,0,0) -- ++(-0.5*\cubex,0,0) -- ++(0.5*\x1,0) coordinate (nw) -- ++(0,-\cubey,0) coordinate (sw) -- ++(\cubex,0,0) node [midway,below,inner sep=0.25em] {\scriptsize$2C$} -- ++(0,\cubey,0) coordinate (ne) -- cycle
    (ne) -- ++(0,0,-\cubez) -- ++(0,-\cubey,0) coordinate (zc) -- ++(0,0,\cubez) node [midway,below,sloped,inner sep=0.25em] {\scriptsize$1$} -- cycle
    (ne) -- ++(0,0,-\cubez) -- ++(-\cubex,0,0) -- ++(0,0,\cubez) -- cycle;

    \draw[fill=Paired-B] let \p1 = (0,0,\cubez) in
    (0,-\cubehy+\cubeufy,0) -- ++(-0.5*\cubex,0,0) -- ++(0.5*\x1,0) coordinate (bn1) -- ++(0,-\cubeufy,0) coordinate (bs1) -- ++(\cubex,0,0) -- ++(0,\cubeufy,0) coordinate (ne) -- cycle
    (ne) -- ++(0,0,-\cubez) -- ++(0,-\cubeufy,0) -- ++(0,0,\cubez) -- cycle;
    
    \draw[fill=Paired-B] let \p1 = (0,0,\cubez) in
    (0,-\cubey+\cubeufy,0) -- ++(-0.5*\cubex,0,0) -- ++(0.5*\x1,0) coordinate (bn2) -- ++(0,-\cubeufy,0) coordinate (bs2) -- ++(\cubex,0,0) -- ++(0,\cubeufy,0) coordinate (ne) -- cycle
    (ne) -- ++(0,0,-\cubez) -- ++(0,-\cubeufy,0) -- ++(0,0,\cubez) -- cycle;

    \draw[lbrace=0.15] (nw) -- (sw) node[pos=0.15,left=4pt,inner sep=0.25em] {\rotatebox{90}{\scriptsize$4C$}};
    
    \draw[lfbrace=0.5,draw=none] (bn1) -- (bs2) node[midway,left=7pt,inner sep=0.25em] {\rotatebox{90}{\scriptsize$\left\lceil4C\ufj\right\rceil$}};

    \begin{scope}
        \clip (bs1) -- (bn2) -- ++(-1em,0) |- (bs1) -- cycle;
        \draw[lfbrace=0.5,densely dashed] (bn1) -- (bs2);
    \end{scope}

    \begin{scope}[even odd rule]
        \clip
        (bn1) -- (bs2) -- ++(-1em,0) |- (bn1) -- cycle
        (bs1) -- (bn2) -- ++(-1em,0) |- (bs1) -- cycle;
        \draw[lfbrace=0.5] (bn1) -- (bs2);
    \end{scope}
\end{scope}

\node [fill=none, right=3em of trconv, minimum width=4.5em, minimum height=2.5em] (convcanv) {};
\pgfmathsetmacro{\cubex}{4}
\pgfmathsetmacro{\cubey}{2}
\pgfmathsetmacro{\cubez}{1.2}
\pgfmathsetmacro{\cubefr}{0.75}
\pgfmathsetmacro{\cubefrr}{1-\cubefr}
\begin{scope}[shift={($(convcanv.north) + (0,-0.5em)$)},line join=round]
    \draw[fill=Paired-D] let \p1 = (0,0,\cubez) in
    (0,0,0) -- ++(-0.5*\cubex,0,0) -- ++(0.5*\x1,0) -- ++(0,-\cubey,0) node[midway,left,inner sep=0.25em] {\rotatebox{90}{\scriptsize$C$}} coordinate (sw) -- ++(\cubex,0,0) coordinate (se) -- ++(0,\cubey,0) coordinate (ne) -- cycle
    (ne) -- ++(0,0,-\cubez) -- ++(0,-\cubey,0) coordinate (zc) -- ++(0,0,\cubez) node [midway,below,sloped,inner sep=0.25em] {\scriptsize$K$} -- cycle
    (ne) -- ++(0,0,-\cubez) -- ++(-\cubex,0,0) -- ++(0,0,\cubez) -- cycle;
    
    \draw[fill=Paired-C] let \p1 = (0,0,\cubez) in
    (0,0,0) -- ++(-\cubefrr*\cubex,0,0) -- ++(0.5*\x1,0) -- ++(0,-\cubey,0) -- ++(\cubefr*\cubex,0,0) -- ++(0,\cubey,0) coordinate (ne) -- cycle
    (ne) -- ++(0,0,-\cubez) -- ++(0,-\cubey,0) -- ++(0,0,\cubez) -- cycle
    (ne) -- ++(0,0,-\cubez) -- ++(-\cubefr*\cubex,0,0) -- ++(0,0,\cubez) -- cycle;

    \draw[lbrace=0.75] (sw) -- (se) node[pos=0.75,below=4pt,inner sep=0.25em] {\scriptsize$2C$};
    
    \draw[lfbrace=0.5] (sw) -- ++(\cubefrr*\cubex,0,0) node[midway,below=7pt,inner sep=0.25em] {\scriptsize$\left\lceil2C\ufj\right\rceil$};
\end{scope}


\draw [ar] (in1) |- (sum);
\draw [ar] (in2) |- (sum);
\draw [ar] (sum) -- (pwconv)  node[arlbl,pos=0.4] {\scriptsize$\left[2C, \frac{T}{S}\right]$};
\draw [ar] (pwconv) -- (glu)   node[arlbl] {\scriptsize$\left[4C\ufj, \frac{T}{S}\right]$};
\draw [ar] (glu) -- (trconv) node[arlbl] {\scriptsize$\left[2C\ufj, \frac{T}{S}\right]$};
\draw [ar] (trconv) -- (relu);
\draw [ar] (relu) -- (out)   node[arlbl] {\scriptsize$\left[C, T\right]$};

\pgfresetboundingbox \useasboundingbox[draw=none] (tensorlbl.north east) -| (glu.west) |- (out) -| (tensorlbl.north east);

\end{tikzpicture}
        }
        \caption{Decoder block $\mathcal{D}_i$}
        \label{fig:slim_block_dec}
    \end{subfigure}
    \caption{Slimmable blocks with weight tensors; width, height, and depth correspond to input channels, output channels, and kernel size, respectively; $C = 2^{i-2}H$ for $i \ge 2$ is the current hidden size.}
    \label{fig:slim_block}
\end{figure}
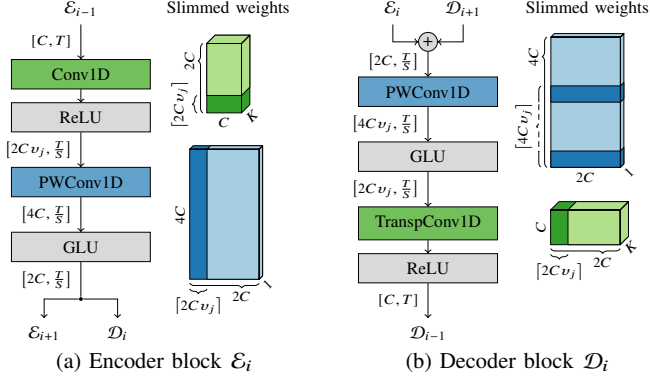

\section{Related Work}
\label{sec:related}

\begin{description}[
    font=\bfseries,
    leftmargin=0pt,
    parsep=\parsep,
    listparindent=\parindent,
    labelwidth=0em,
    itemindent=1em,
    labelsep=1em,
    align=left,
    itemsep=\parsep,
]

\item[Efficient speech enhancement]
Over the years, several deep learning models for SE have demonstrated real-time capabilities.
These efficient SE models may integrate architectural elements such as multi-path processing~\cite{shankar_real-time_2020,valin_perceptually-motivated_2020,yu_pfrnet_2022} or encoder-decoder topology with skip connections~\cite{stoller_wave-u-net_2018,braun_towards_2021,sach_effcrn_2023,rong_gtcrn_2024} based on U-Net~\cite{ronneberger_u-net_2015}.
Some of these architectures, often based on recurrent networks~\cite{braun_data_2020}, have been specifically optimized for embedded devices~\cite{fedorov_tinylstms_2020,rusci_accelerating_2023}.

\item[Dynamic networks]
DynNNs can adapt their structure based on the input~\cite{han_dynamic_2022}.
Early-exiting models control how many layers are used~\cite{scardapane_why_2020} while recursive networks can vary how many times they are executed~\cite{guo_dynamic_2019}.
Slimming~\cite{yu_slimmable_2019,li_dynamic_2021} or channel pruning~\cite{herrmann_channel_2020,gao_dynamic_2019} affect the model width, i.e., the number of active nodes in a given layer.
Finally, dynamic routing and the mixture of experts (MoE) can switch between different weights~\cite{fedus_switch_2022} or entire subnetworks~\cite{shazeer_outrageously_2017}.

\item[DynNNs for audio]
Several DynNN techniques have been applied to SE and source separation, including early-exiting~\cite{chen_dont_2021,li_learning_2021,kim_bloom-net_2022,miccini_dynamic_2023}, recursive networks~\cite{bralios_latent_2023,kim_scalable_2024}, slimming~\cite{elminshawi_slim-tasnet_2023,elminshawi_dynamic_2024}, channel pruning~\cite{miccini_scalable_2025} and MoE~\cite{sivaraman_sparse_2020,zezario_speech_2021}.
A recent work on source separation applied dynamic slimming to the feed-forward layers of transformer blocks~\cite{elminshawi_dynamic_2024}. 
In contrast, we slim a larger part of the network and employ a single router to reduce overhead. 
We also propose an improved training scheme based on the \textit{Gumbel-softmax trick}~\cite{maddison_concrete_2017,jang_categorical_2017} and demonstrate its effectiveness with more UFs, resulting in a system that offers greater computational savings and more versatility.

\end{description}

\begin{figure*}[t]
    \centering
    \centerline{\includegraphics[width=0.95\textwidth]{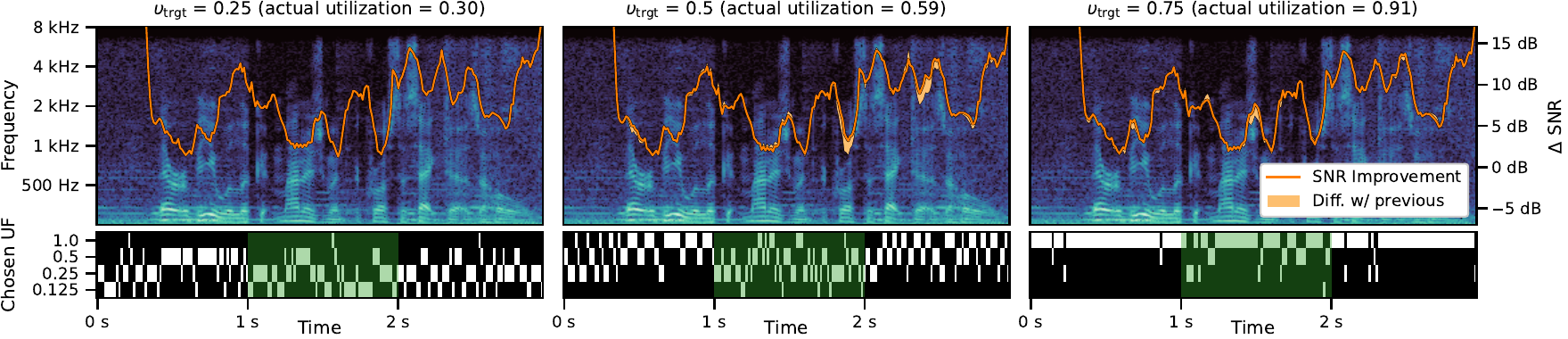}}
    \caption{Example inference for models trained with $\uftrgt \in \left\{ 0.25, 0.5, 0.75 \right\}$, showing input spectrogram along with improvement in instantaneous SNR (orange line, with yellow regions marking difference with preceding model on the left) and chosen UFs over time. In the middle portion of the sample (in green), we deliberately reduce the noise level by \qty{10}{\decibel} to show how each model reacts by decreasing its average utilization.}
    \label{fig:example}
    \vspace{-0.2em}
\end{figure*}

\section{Dynamic Slimmable DEMUCS}
\label{sec:dynslim_demus}

\subsection{DEMUCS for Speech Enhancement}
\label{ssec:backbone}

Given a single-channel noisy audio signal $x$ of length $T$, composed of clean speech $s$ and additive noise $n$, we compute an estimate of the speech $\hat{s}$ using a function $\mathcal{M}$ such that $\mathcal{M}(x) = \hat{s}$.
Here, our choice of $\mathcal{M}$ falls on the DEMUCS architecture, initially developed for music source separation~\cite{defossez_demucs_2019} and subsequently adapted for real-time monaural speech enhancement~\cite{defossez_real_2020}.
This architecture consists of an encoder, a bottleneck, and a decoder, as shown in \cref{fig:dynslim_demucs}.

The encoder is composed of $L$ blocks, each denoted as $\mathcal{E}_i$, featuring 1D convolution with kernel size $K$ and stride $S$, ReLU activation, pointwise 1D convolution, and gated linear unit (GLU)~\cite{dauphin_language_2017}.
Each encoder block downsamples the signal by a factor $S$ and doubles its channels, starting from an initial value $H$.
Following a symmetric structure, decoder blocks $\mathcal{D}_i$ feature pointwise 1D convolution, GLU, transposed 1D convolution and --- except for the last block --- ReLU activation.
In order to handle waveform input and output, both $\mathcal{E}_1$ and $\mathcal{D}_1$ have a single input and output channel, respectively.

The bottleneck performs sequence modeling on a high-dimensional space comprising $2^{L-1}H$ features.
The original DEMUCS employs two long short-term memory (LSTM) units spanning the entire feature space.
To reduce the number of trainable parameters and operations, we split the feature space into $M$ equally-sized blocks, assign each block to a separate gated recurrent unit (GRU), and concatenate the resulting output activations into a feature vector with the same dimensionality as the bottleneck input.
This is equivalent to employing a single GRU where non-zero values occur only in blocks along the weight matrix diagonals, allowing full connectivity within a block and no connectivity across groups~\cite{keirsbilck_rethinking_2019,braun_towards_2021}.
Lastly, we employ skip connections, as popularized by the U-Net architecture~\cite{ronneberger_u-net_2015}, between encoder and decoder blocks, and operate on a version of the input signal $x$ that is upsampled by a factor $U$ and subsequently downsampled.

\subsection{Slimmable Encoder-Decoder Blocks}
\label{ssec:slim_blocks}

At the core of our blocks, we use slimmable 1D convolution. 
Each convolutional layer maps a utilization factor, denoted as $\ufj \in \mathbb{U}$, to a specific portion of convolutional filters. 
We define $\mathbb{U} = \smash{\left\{ \ufj \right\}}^J_{j=1}$ as the discrete set of available UFs of size $J$. 
These factors $\ufj$ are fractional values ranging from partial utilization, i.e., $\upsilon_j < 1$, to full utilization, i.e., $\upsilon_J = 1$, controlling the proportion of active weights in the layer. 
Thus, $\mathbb{U}$ can be treated as an architectural hyperparameter.
During inference, the slimmable layers receive a UF as additional input, determining the subset of active filters. 
Given the convolutional weight tensor $W \in \mathbb{R}^{C_{\text{out}} \times C_{\text{in}} \times K}$, slimming can be applied along the output channels, the input channels, or both. 
This process involves selecting the first $\left\lceil C_{\text{out}} \ufj \right\rceil$ output channels and/or $\left\lceil C_{\text{in}} \ufj \right\rceil$ input channels, depending on the slimming direction.

As shown in \cref{fig:slim_block_enc}, we slim the first convolution in the encoder block (green) along its output and the pointwise convolution (blue) along its input.
Thus, only the block's internal feature space is slimmed, while its input and output maintain the same dimensionality regardless of UF.
In this way, we avoid slimming the GLU, which would cause certain features to be treated as either gating or activation depending on the UF.
The decoder block in \cref{fig:slim_block_dec} follows a similar scheme, preserving the dimensionality of its input and output while slimming internally.
However, since the GLU is now applied on the slimmed signal, we employ a slightly different approach for the pointwise convolution (blue).
To comply with the GLU structure, we retrieve half of the required weights from the start of the tensor and the other half from the middle, ensuring that the gating portion of the GLU input is always processed with filters associated with it.

\subsection{Routing sub-network}
\label{ssec:ufpicker}
The routing module $\mathcal{R}$, shown in \cref{fig:dynslim_demucs} (red), is tasked with selecting a UF based on the input.
It comprises 1D convolution, ReLU activation, diagonal GRU, and point-wise convolution.
To limit its computational overhead, we employ diagonal GRUs~\cite{subakan_diagonal_2017}, which can be seen as an extreme case of grouped GRUs where each hidden feature has a separate GRU~\cite{keirsbilck_rethinking_2019}, effectively replacing all matrix multiplications with element-wise products.
Furthermore, we match stride and kernel size in the first convolutional layer, decreasing its time resolution and resulting in the downsampled sequence $T^{\ast} < T$.

This subnet computes a score matrix $r \in \mathbb{R}^{J \times T^{\ast}}$, such that the distribution of UFs for each input frame is given by $\softmax_{j}(r)$.
Since only one UF out of the $J$ available ones is used at any given time, we must turn the aforementioned distribution into discrete decisions.
The most straightforward approach involves selecting the UF with the highest probability using the $\argmax$ operator.
However, this operation is not differentiable, preventing end-to-end training.
Furthermore, it results in selections that do not reflect the distribution estimated by $\mathcal{R}$, causing low-probability UFs to never be chosen.
To address both issues, we apply the Gumbel-softmax trick proposed in \cite{maddison_concrete_2017,jang_categorical_2017}: during training, we sample from the distribution by adding Gumbel noise to $r$, ensuring exploration of all available UFs.
Differentiation is thus supported through the following relaxation:
\begin{equation}
    \mathcal{R}(x) =\begin{cases}
        \argmax_{j} (r + G) & \text{Forward pass}\\
        \softmax_{j} (r + G) & \text{Backward pass}
    \end{cases}
\end{equation}
where $G$ is a matrix of i.i.d. Gumbel noise samples.
This allows the routing subnet to provide discrete UFs while backpropagating an estimated gradient.
During inference, we omit the noise, i.e., $G = 0$.

\subsection{Training Strategy}
\label{ssec:training}

The training process comprises two stages: firstly, we pre-train the slimmable DEMUCS architecture to ensure proficiency in the denoising task for all UFs.
Subsequently, we implant the routing subnet $\mathcal{R}$ and train it end-to-end together with the slimmable backbone.

During the first stage, we predict a signal $\hat{s}_j$ using each $\mathcal{M}_j$, i.e., the model running at utilization $\ufj$.
We then compute a denoising loss $\mathcal{L}_\text{SE}$ for each of the signals $\hat{s}_j$, and minimize their sum:
\begin{equation}
    \mathcal{L}_\text{Slim} = \sum_{j} \mathcal{L}_\text{SE} \left( s, \hat{s}_j \right) \qquad \hat{s}_j = \mathcal{M}_j \left( x \right)
    \label{eq:loss_slim}
\end{equation}

We employ the loss described in \cite{braun_data_2020}, which we found empirically to yield better results than the one originally used by DEMUCS.
It computes the squared distances between predicted and target signals of their complex and magnitude STFTs (with $l$ and $f$ indexing time and frequency bins), both compressed by $c$ and mixed by a factor $\alpha$:
\begin{equation}
    \mathcal{L}_\text{SE} = \alpha \sum_{l, f} \left| |S|^c e^{j \angle_{\scaleto{S\rule{0ex}{1.5ex}}{0.5em}}} - |\hat{S}|^c e^{j \angle_{\scaleto{\hat{S}\rule{0ex}{1.5ex}}{0.5em}}} \right|^{\scriptscriptstyle 2} + (1 - \alpha) \sum_{l, f} \left| |S|^c - |\hat{S}|^c\right|^{\scriptscriptstyle 2}
    \label{eq:loss_se}
\end{equation}

In the second stage, we minimize an end-to-end loss comprising the following regularization terms ($\beta$ and $\gamma$ are weighting coefficients):
\begin{equation}
    \label{eq:loss_dynslim}
    \mathcal{L}_\text{DynSlim} = \mathcal{L}_\text{SE}(s, \hat{s}) + \beta \mathcal{L}_\text{Eff} + \gamma \mathcal{L}_\text{Bal}
\end{equation}

The estimated clean speech $\hat{s}$ is computed by combining the output from each UF using using $\vmathbb{1}_{\mathcal{R}(x) = j}$, an indicator function that is \num{1} when UF $\ufj$ is selected and \num{0} otherwise, as gating signal.
Since $\mathcal{R}(x)$ operates at a lower resolution $T^{\ast}$, we first resample its output to match the bottleneck resolution $\frac{UT}{S^L}$ using nearest-neighbor interpolation, followed by zero-order hold to obtain the original input length $T$:

\begin{equation}
    \hat{s} = \sum_j \mathcal{M}_j(x) \cdot \upsample_{T^{\ast} \rightarrow T}\left( \vmathbb{1}_{\mathcal{R}(x) = j} \right)
\end{equation}

The loss term $\mathcal{L}_\text{Eff}$ ensures the learned model achieves the defined level of efficiency by keeping the overall utilization close to a target value $\uftrgt$.
Meanwhile, the balancing term $\mathcal{L}_\text{Bal}$, based on \cite{fedus_switch_2022}, prevents the model from favoring only a few UFs.
These are:
\begin{equation}
    \label{eq:losses_dynslim}
    \mathcal{L}_\text{Eff} = \bigg( \sum_j \Big( \tilde{\Upsilon}_j \ufj \Big) - \uftrgt \bigg)^2, \quad  \mathcal{L}_\text{Bal} = \frac{1}{J-1} \bigg( J \sum_j {\tilde{\Upsilon}_j}^2 - 1 \bigg)
\end{equation}
where $\tilde{\Upsilon} \in \mathbb{R}^{J}$ represents the relative occurrence of each UF across a training batch such that $\tilde{\Upsilon}_j = \mathbb{E} \left( \vmathbb{1}_{\mathcal{R}(x) = j} \right)$.
We use this because, although we want to favor smaller UFs, we still expect the model to employ bigger UFs on challenging input data, as long as the target efficiency is maintained globally.


\begin{figure*}[t]
    \centering
    \centerline{\includegraphics[width=0.95\textwidth]{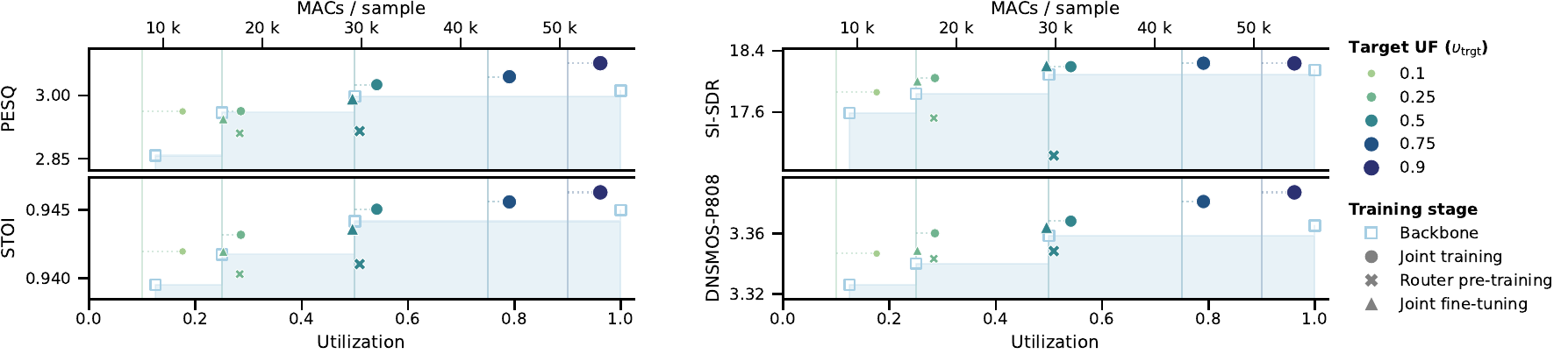}}
    \caption{Pareto fronts given by utilization/MACs (x-axes) vs. speech quality metrics (y-axis) for static slimmable backbone (light blue empty squares) and dynamic models trained on a range of $\uftrgt$; the dotted horizontal lines show the distance between each $\uftrgt$ (solid vertical lines) and actual average utilization. For $\uftrgt \in \left\{0.25, 0.5 \right\}$, we include results from the alternative training schedule described in \cref{itm:exp_training}.}
    \label{fig:pareto}
\end{figure*}

\begin{figure}[t]
    \centering
    \centerline{\includegraphics[width=0.95\columnwidth]{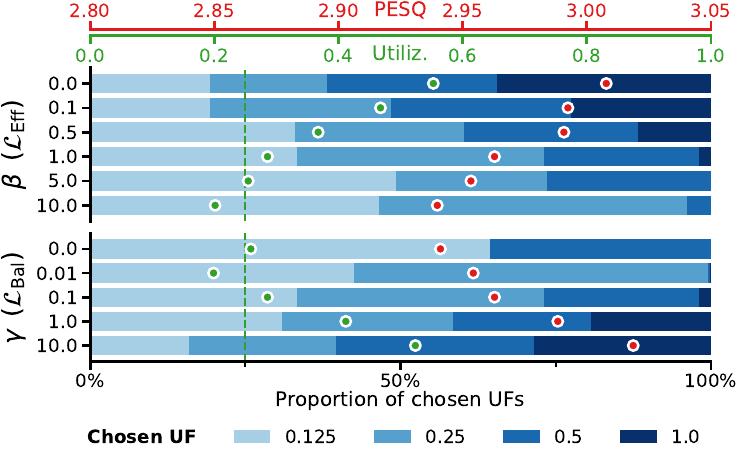}}
    \caption{For models trained with different regularization coefficients (one per row), we show the relative occurrence of each UF (i.e. $\tilde{\Upsilon}_j$; blue stacked bars) along with their PESQ and avg. utilization (red and green dots, uppermost x-axes); dashed line showing $\uftrgt=0.25$.\vspace{-0.1em}}
    \label{fig:weights_both}
\end{figure}

\section{Experimental Setup}
\label{sec:exp_setup}

\begin{description}[
    font=\bfseries,
    leftmargin=0pt,
    parsep=\parsep,
    listparindent=\parindent,
    labelwidth=0em,
    itemindent=1em,
    labelsep=1em,
    align=left,
    itemsep=\parsep,
]

\item[Data] 
We perform our experiments on the popular VoiceBank+\allowbreak DEMAND dataset~\cite{valentini-botinhao_investigating_2016}.
The training split features \num{11572} utterances from \num{28} speakers, corresponding to \num{\sim 10} hours of data, with background noise mixed at SNRs ranging from \qty{0}{\decibel} to \qty{15}{\decibel}.
We use the utterances from two random speakers as a validation set.
The test split contains \num{824} utterances from \num{2} unseen speakers mixed with unseen noise at SNRs between \qty{2.5}{\decibel} and \qty{17.5}{\decibel}.
The data is downsampled to \qty{16}{\kilo\hertz}.
During training, we randomly extract segments of \qty{4}{\second} and augment them based on the DEMUCS training recipe~\cite{defossez_real_2020}.

\item[Model]
Similarly to the causal variant in the original paper, our DEMUCS backbone is parameterized by $L=5$, $K=8$, $S=4$, and $U=4$.
We decrease the initial hidden channels $H$ to \num{32} and employ the grouped GRU method described in \cref{ssec:backbone} with $M=4$.
For slimming, we set $\mathbb{U}=\left\{0.125, 0.25, 0.5, 1.0\right\}$ and thus $J=4$.
The layers in the subnet $\mathcal{R}$ have kernel size of \num{256} and \num{64} hidden features/channels, introducing a negligible \qty{0.06}{\percent} overhead.

\item[Training]
\phantomsection\label{itm:exp_training}
During the first stage, we pre-train the slimmable backbone for up to \num{400} epochs on batches of \num{32} elements, using a learning rate of \num{1e-3} that is halved every \num{15} epochs without improvement on the validation set.
In the second stage, we transfer the pre-trained backbone weights, randomly initialize the router, and train both for up to \num{250} epochs on batches of \num{16} elements with a learning rate of \num{1e-3} decaying by a factor of \num{0.99} per epoch.
We also test a variant where the router is pre-trained with the backbone frozen, followed by joint fine-tuning at a learning rate of \num{1e-4} (\ding{54} and \ding{115} in \cref{fig:pareto}).
We use $c = 0.3$, $\alpha = 0.3$ for the SE loss \cref{eq:loss_se} (as in the original paper), and set $\beta = 1.0$, $\gamma = 0.1$ for the end-to-end loss \cref{eq:loss_dynslim}, based on observations in \cref{fig:weights_both}.
In all cases, we use the Adam optimizer and enable early stopping with a patience of \num{35} epochs.

\item[Metrics]
We assess speech quality with the perceptual evaluation of speech quality (PESQ)~\cite{rix_perceptual_2001}, the scale-invariant signal-to-distortion ratio (SI-SDR)~\cite{le_roux_sdr_2019}, the short-time objective intelligibility (STOI)~\cite{taal_short-time_2010}, and mean opinion scores predicted using DNSMOS P.808~\cite{reddy_dnsmos_2021}.
For computational efficiency, we estimate multiply-accumulate operations (MACs) per input sample using \texttt{fvcore}\footnote{\url{https://github.com/facebookresearch/fvcore/blob/main/docs/flop_count.md}}.
\end{description}

\section{Results}
\label{sec:results}
\cref{fig:pareto} presents Pareto fronts comparing dynamically-slimmable models trained with different $\uftrgt$ against the static slimmable backbone.
The dynamic models consistently achieve Pareto efficiency across all metrics, demonstrating improved trade-offs between performance and efficiency.
The proposed method is more advantageous when the system is asked to reduce the complexity. 
For example, when training with $\uftrgt = 0.1$ (smallest dot), performance is on par with or better than the backbone with static \num{0.25} utilization (empty square on line at \num{0.25}) for all metrics. 
With regularization from $\mathcal{L}_\text{Eff}$, the actual utilization is \num{0.176} on average, resulting in a \qty{29}{\percent} reduction in MACs when compared to the static backbone at UF = \num{0.25}. 
Similarly, when training for $\uftrgt = 0.9$ (largest dot), the actual utilization averages \num{0.962}, resulting in a \qty{4}{\percent} MACs reduction while improving PESQ by \qty{2.2}{\percent} and SI-SDR by \qty{0.5}{\percent} when compared to the full static backbone (i.e., UF = \num{1}). 
Notably, splitting the second training stage into router pre-training followed by joint fine-tuning achieves better adherence to utilization targets but proves detrimental to speech quality.
 
The impact of the regularization coefficients used in \cref{eq:loss_dynslim} is shown in \cref{fig:weights_both}.
Expectedly, higher $\beta$ values improve target adherence, which leads to narrower distributions concentrated around small UFs, whereas lower values yield a more diverse selection of UFs, improving PESQ scores despite exceeding $\uftrgt$.
The choice $\gamma$ has an analogous yet inverse effect: in both cases, a clear correlation between UF uniformity, average utilization, and speech quality is observed.

Lastly, \cref{fig:dynslim_jointplot} demonstrates how a dynamic model adapts its capacity based on input difficulty, indicated there by the initial SI-SDR of noisy inputs. 
We observe that easier inputs (higher SI-SDR) are processed using lower UFs on average while more capacity is allocated for noisier inputs, leading to larger SI-SDR improvements. 
Intuitively, this means that resources are spent predominantly on challenging inputs requiring more enhancement while realizing savings on easier ones, reflecting the merit of the proposed architecture.
This is further exemplified in \cref{fig:example}, where input regions featuring lower noise levels or less speech content trigger the selection of lower UFs.

\section{Conclusion}
\label{sec:conclusion}
We introduced dynamic slimming to SE, extending the DEMUCS architecture with slimmable blocks and a lightweight routing network. 
The proposed system dynamically scales to match the complexity of incoming audio, improving quality on challenging inputs while saving resources on easier ones. 
Our method demonstrates Pareto-optimal behavior across speech quality metrics, offering better trade-offs than static slimming.
This highlights the practical value of input adaptiveness and computational scalability, as  showcased by our dynamically-slimmable networks, in real-time, resource-constrained settings.
We plan to extend this approach to encompass the recurrent bottleneck, leverage the model's internal representations for routing, and verify its applicability on alternative SE architectures.

\bibliographystyle{IEEEtran2}
\bibliography{IEEEabrv,references,ctrl}

\end{document}